# Very unusual operation of the electron accelerator above Aragats mountain in Armenia a day after the earthquake in Turkey and Syria

A. Chilingarian , G. Hovsepyan, D. Aslanyan ,T. Karapetyan, B. Sargsyan

A. Alikhanyan National Lab (Yerevan Physics Institute), Yerevan 0036, Armenia

During the multiyear monitoring of particle fluxes and near-surface electric field (NSEF) on the Aragats research station, no runaway relativistic electron avalanches have been observed in the January-February months. The large peaks and energies of TGE particles originating from the electron-gamma avalanches in the thundercloud are observed in the Spring-Autumn months when the electric field inside the cloud often exceeds the runaway threshold strength. On February 7, 2023, suddenly, all particle detectors registered 3 TGEs within 10 hours without any unusual local weather conditions. We consider these TGEs as indirect evidence of the influence of strong earthquakes on the previous day on the ionosphere and via the ionosphere on the structure and strength of the intracloud electric field above Aragats mountain. Recovered energy spectra of TGE electrons and gamma rays prove nearly one hour the strength of the electric field above Aragats at the heights 3300-5300 m comprises ≈2.1 kV/cm.

## 1. Introduction

An electron accelerator in thunderclouds sends copious electrons, gamma rays, and rarely neutrons to the earth. During the last decade, on Mt. Aragats in Armenia, Mt. Lomnicky Stit in Slovakia, and Mt. Musala in Bulgaria registered nearly a thousand so-called thunderstorm ground enhancements (TGEs, [1.2]), abrupt enhancements of cosmic ray fluxes, the strongest of which surpass the background ≈200 times [3]. Acceleration and multiplication of free atmospheric electrons are possible if the intracloud electric field is larger than the critical value, specific for the particular air density [4-6]. The theory of relativistic runaway avalanche was given in [7], and the model of a dipole accelerated electrons downward – in [8]. A vast amount of fully described TGE events are available from Mendeley datasets [9-11] and the database of the Cosmic Ray Division of Yerevan Physics Institute [12]. The particle detectors and data analysis methods are described in [13-15].

Usually, TGEs are registered in the Spring-Autumn seasons (April-October on Aragats) when the outside temperature is not lower than -5C. The near-surface electric field during thunderstorms on Aragats reaches -20 – -30 kV/ m, and the intracloud electric field exceeds 2.1 kV/cm, sometimes prolonging almost to the earth's surface (3200 m asl). In the Winter season (November-February), disturbances of NSEF are much more minor, and the intracloud electric field does not reach the critical energy to unleash electron-photon avalanches. During 10 years of coherent monitoring of cosmic ray fluxes, electric fields, and meteorological parameters, we registered only 3 small TGEs in the Winter season and no one in January-February (see discussion in [16]). Fig.1 shows that most TGE occurred in the -3C +3C interval in April-October.

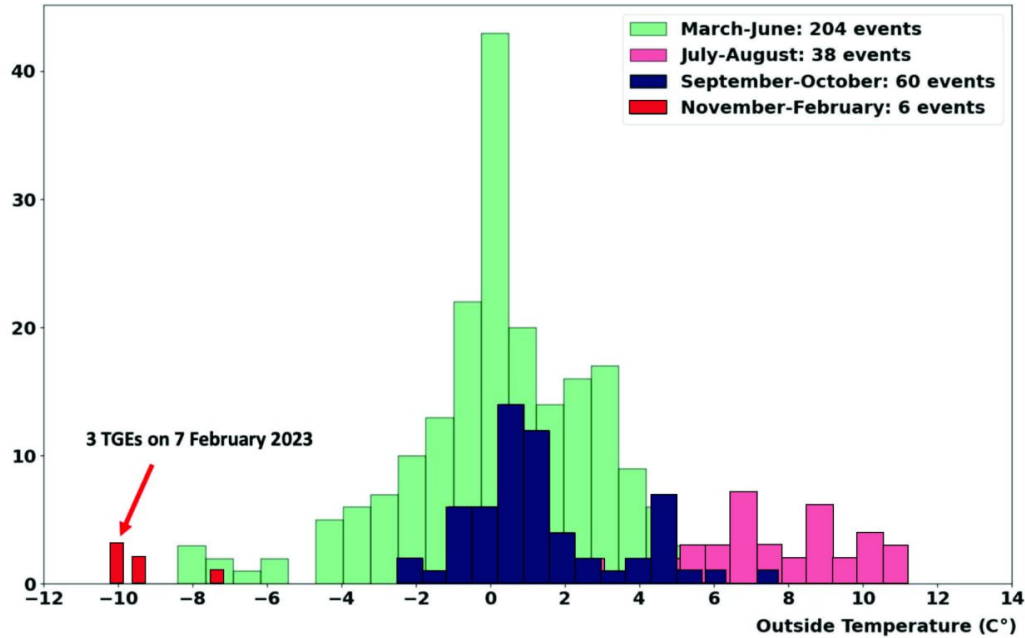

**Figure 1. The season-dependent histogram of TGE occurrences dependent on the outside temperature**

Suddenly, on February 7, we observed 3 TGE events in 10 hours at temperatures ≈-10C. Proceeding from this very improbable series of TGEs, we present a detailed description of the observations on Aragats and discuss their possible relation to the largest earthquake in Turkey on 6 February.

2. **Observation of enhanced particle fluxes on Mt. Aragats on 7 February 2023**

In Fig.2, we show the MSEF disturbances, the time series of the count rate of a 3-cm thick plastic scintillator, and the outside temperature. The NSEF disturbances during the first largest TGE ranged from -15 kV/m to +9 kV/m; the Outside atmosphere was approximately -10C; TGE significance by STAND3 upper detector (coincidence "1000" was ≈ 10σ).

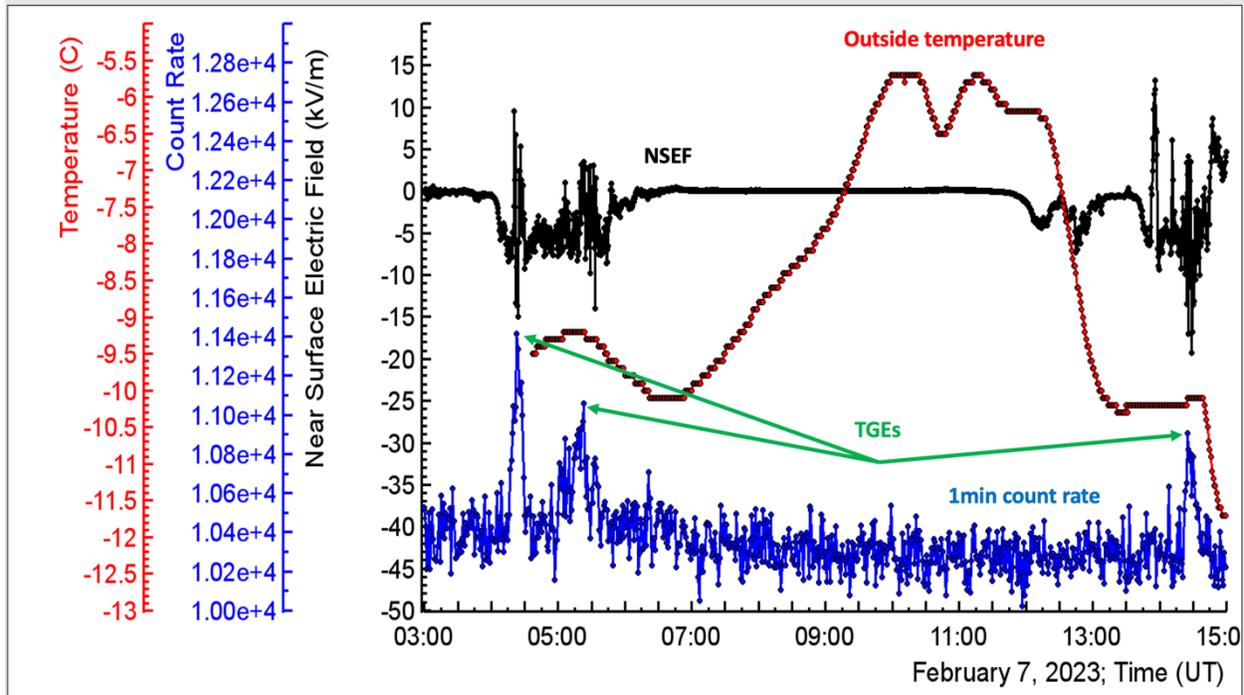

**Figure 2.** Black - disturbances of NSEF, measured by the electric mill EFM-100 produced by Boltek firm; blue- time series of count rate of upper scintillator of STAND3 detector ("1000" coincidence, signal only in the upper scintillator of the stacked assembly); red – outside temperature measured by Davis weather station. By green arrows, we show 3 TGEs that occurred within ≈10 hours.

The relative humidity was 92%, atmospheric pressure was 673.6 millibar. The wind declines during the first and third TGEs and reaches shortly five m/s only during the second TGE, see Table 1. Table 1 shows 6 TGE events that occurred in the Wintertime. One can notice that in 10 years, no TGE occurred in January and February besides 7 February 2023. The significance of the TGE is measured in the number of standard deviations above the mean value (Nσ), measured on fair weather before the TGE. The peak value was determined from the time series of "1000" coincidence of the STAND3 detector; signal only in the upper scintillator from 4 stacked one over another. NSEF was measured by EFM 100 electric mill produced by BOLTEK company, outside temperature, and wind speed – by DAVIS weather station.

**Table 1. Winter TGE events occurred in 2013-2023**

| Date | Duration (minutes) | Significance (Nσ) | Outside Temp (C°) | Wind speed (m/sec) | NSEF (kV/m) |
|---|---|---|---|---|---|
| November 30, 2017 | 3:33 – 3:48 | 45 | -7 | 0 | -5 – +1 |
| November 10, 2021 | 0:48 – 1:13 | 6 | -6 | 0 | -13 – +1.7 |
| December 21, 2021 | 12:00 – 12:11 | 7.5 | -10.1 | 5-6 | -14 – +9 |
| February 7, 2023 | 4:15 – 4:31 | 10 | -9,5 | 0 | -15 – +9 |
| February 7, 2023 | 5:11 – 5:28 | 7 | -9,3 | 1-4 | -14 – +3.5 |
| February 7, 2023 | 14:21 – 14:39 | 9 | -10,2 | 0 | -19 – +13 |

Figure 3 shows differential energy spectra of TGE electrons and gamma rays recovered from energy release histograms measured by a 60 cm thick scintillator of the ASNT detector (see description of detector and energy spectra recovering method in [13]). Only for these 2 consequent minutes was it possible to reliably recover the electron energy spectrum. Electrons lose energy to ionization very fast after leaving an accelerating electric field; energy losses of gamma rays are much smaller. Thus, if particles leave the field above 150 m from the ground, TGEs comprises gamma rays only. The presence of electrons in the flux is also confirmed by the "1110" coincidence of the STAND3 detector (signals in 3 scintillators, energy threshold 30MeV). Thus, the electric field during the first TGE reaches 2.1 kV/m (the critical energy on 3300 m asl) at altitudes from 5300 m to 3300 m. It is unprecedented for January-February and was never measured in the previous 10 years.

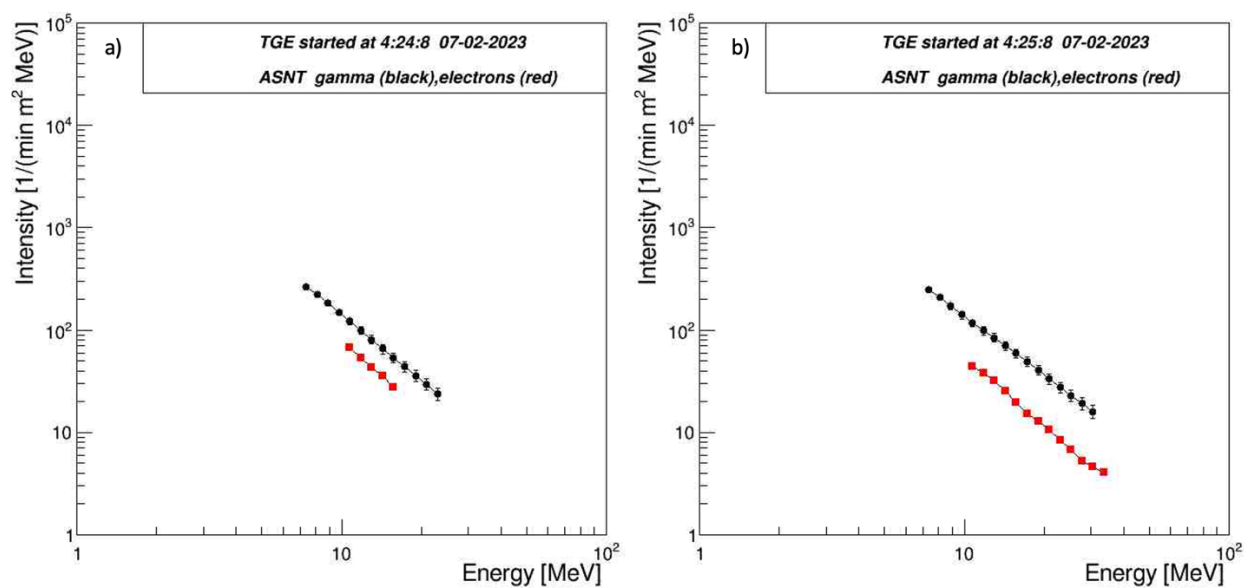

**Figure 3. The differential energy spectra of TGE electrons (red) and gamma rays (black) recovered from energy release histograms measured by the ASNT spectrometer.**

In Table 2, we show the fit parameters of both spectra, the fluxes (integral spectra), and the maximum values of electrons and gamma rays.

**Table 2. Spectral indices, fluxes, maximal energies of TGE particle fluxes, and electron to gamma ray ratio for energies above 10 MeV.**

| 7 February 2023 | Spectral indices | | Integral spectra >10MeV | | Ne/Nγ >10MeV | Max. energy | |
|---|---|---|---|---|---|---|---|
| | elect. | gamma | elect. | gamma | | elect. | gamma |
| 04:24 | 2.19 | 1.79 | 616 | 2176 | 0.28 | 15 | 25 |
| 04:25 | 2.26 | 1.93 | 436 | 1453 | 0.3 | 33.7 | 23 |

## 3. Discussion and conclusions

A very unusual sequence of TGEs was measured on Aragats on the second day after the M7.8 and M7.5 earthquakes, with combined seismic energy released of $4.4 \times 10^{16}$ joules. The earthquakes were characterized by exceptionally long ruptures, with maximum displacements of around 5 and 8 meters for the southern and north arms [17]. Thus, we can expect the strongest ionospheric responses to earthquakes., which occur when the earth's crust reverberates through the atmosphere. The formation of the ever-largest systems of cracks causes disturbances of geophysical fields in the near-surface atmosphere. It possibly enlarges the near-surface electric field (NSEF) strength at large distances from the epicenter. Measurements of the strength of the NSEF, performed at the Geophysical Observatory of Sadovsky Institute of Geosphere Dynamics of the Russian Academy of Sciences, show that during strong earthquakes, there are disturbances of the NSEF at large distances [18]. Possibly, other geophysical anomalies can explain the extreme electric fields in the lower atmosphere above Aragats station. However, the statistics of Winter TGEs show that measured TGEs did not fit the overall statistic of the TGE occurrences (Fig 1). An enormous natural phenomenon is needed to generate observed large electric fields (see Fig 3 and Table 1) in January-February when no thunderstorms occurred for 10 years. Finally, we live on the inner plate of a spherical capacitor, and if it cracks, the outer plate (ionosphere) should feel it. Thus, the most probable explanation of observed TGE events is strong electric fields induced by the disturbed ionosphere.

## ACKNOWLEDGMENTS

We thank the staff of the Aragats Space Environmental Center for the uninterruptable operation of all particle detectors and field meters. The authors acknowledge the support of the Science Committee of the Republic of Armenia (Research Project No. 21AG-1C012) in the modernization of the technical infrastructure of high-altitude stations.